\documentclass[conference]{IEEEtran}

\usepackage{cite}
\usepackage{amsmath,amssymb,amsfonts}
\usepackage{amscd}
\usepackage{epsfig}
\usepackage{psfrag}
\usepackage{rotating}
\usepackage{float}
\usepackage{tikz}
\usetikzlibrary{shapes.geometric}
\usetikzlibrary{arrows.meta}
\usepackage{pgfplots}
\usepackage{mathtools}
\usepackage{algorithmic}
\usepackage{graphicx}
\usepackage{array}
\usepackage{siunitx}
\usepackage{algorithm2e}
\usepackage{float}
\DeclareSIUnit{\belmilliwatt}{Bm}
\DeclareSIUnit{\belsquaremeter}{Bsm}
\DeclareSIUnit{\bit}{bits}
\DeclareMathOperator*{\argmax}{arg\,max}

\graphicspath{ {./figures/} }
\usepackage{textcomp}
\usepackage{xcolor}
\usepackage[nolist,nohyperlinks]{acronym}
\pgfplotsset{compat=1.18}
\long\def\comment#1{}

\newfont{\bbb}{msbm10 scaled 700}

\newfont{\bb}{msbm10 scaled 1100}

\newcommand{\CC}{\mbox{\bb C}}

\newcommand{\RR}{\mbox{\bb R}}

\newcommand{\hv}{{\bf h}}

\newcommand{\qv}{{\bf q}}
\newcommand{\rv}{{\bf r}}
\newcommand{\sv}{{\bf s}}

\newcommand{\wv}{{\bf w}}

\newcommand{\xv}{{\bf x}}
\newcommand{\yv}{{\bf y}}

\newcommand{\Fm}{{\bf F}}
\newcommand{\Gm}{{\bf G}}
\newcommand{\Hm}{{\bf H}}
\newcommand{\Id}{{\bf I}}

\newcommand{\Sm}{{\bf S}}

\newcommand{\Wm}{{\bf W}}

\newcommand{\Xm}{{\bf X}}
\newcommand{\Ym}{{\bf Y}}

\newcommand{\Xc}{{\cal X}}

\newcommand{\Deltam}{\hbox{\boldmath$\Delta$}}

\newcommand{\Pim}{\hbox{\boldmath$\Pi$}}
\newcommand{\Psim}{\hbox{\boldmath$\Psi$}}

\newcommand{\diag}{{\hbox{diag}}}

\newcommand{\herm}{{\sf H}}
\newcommand{\transp}{{\sf T}}

\newcommand{\norm}[1]{\left|\left| #1 \right|\right|}

\usepackage{hyperref}
\hypersetup{
    bookmarks=false,         
    unicode=false,          
    pdftoolbar=true,        
    pdfmenubar=true,        
    pdffitwindow=false,     
    pdfstartview={FitH},    
    pdfnewwindow=true,      
    colorlinks=true,       
    linkcolor=red,          
    citecolor=green,        
    filecolor=magenta,      
    urlcolor=cyan           
}

\def\BibTeX{{\rm B\kern-.05em{\sc i\kern-.025em b}\kern-.08em
    T\kern-.1667em\lower.7ex\hbox{E}\kern-.125emX}}
\begin{document}

\title{Channel Estimation in Uplink Multi-User Scenario using OTFS Modulation\\
\thanks{The code is available on: \url{https://github.com/LorenzoZaniboni/OTFS_OMP.git}}
}

	\author{\IEEEauthorblockN{
            Yatish Pachigolla\textsuperscript{\textsection},
            Lorenzo Zaniboni\textsuperscript{\textsection},
			Mahdi Mahvari\textsuperscript{\textsection}}
		    \IEEEauthorblockA{\\
		    Technical University of Munich, Germany\\
			Emails: \{yatish.pachigolla,
        lorenzo.zaniboni,  mahdi.mahvari\}@tum.de}}

\maketitle
\begingroup\renewcommand\thefootnote{\textsection}
\footnotetext{All authors have contributed equally.}
\endgroup

\begin{acronym}
    \acro{OFDM}{orthogonal frequency-division multiplexing}
    \acro{Tx}{transmitter}
    \acro{Rx}{receiver}
    \acro{IRS}{intelligent reflecting surface}
	\acro{AWGN}{additive white Gaussian noise}
	\acro{MIMO}{multiple-input multiple-output}
	\acro{ULA}{uniform linear array}
	\acro{CRLB}{Cram\'er-Rao lower bound}
	\acro{SNR}{signal-to-noise ratio}
	\acro{mmWave}{millimeter wave}
	\acro{ML}{maximum likelihood}
	\acro{BS}{base station}
	\acro{UE}{user equipment}
	\acro{HDA}{hybrid digital-analog} 
	\acro{ISAC}{integrated sensing and communication}
	\acro{RF}{radio frequency}
	\acro{BA}{beam alignment}
	\acro{AoA}{angle of arrival}
	\acro{AoD}{angle of departure}
	\acro{LOS}{line-of-sight}
	\acro{CP}{cyclic prefix}
	\acro{ISI}{inter-symbol interference}
	\acro{BF}{beamforming}
	\acro{DFT}{discrete Fourier transform}
	\acro{RCS}{radar cross-section}
	\acro{MMLE}{multi-slot maximum likelihood estimation}
	\acro{HIRS}{hybrid-intelligent reflective surface}  
	\acro{DL}{downlink}
	\acro{UL}{uplink}
	\acro{RMSE}{root mean square error}
    \acro{AP}{access point}
    \acro{TDD}{time-division duplex}
    \acro{OTFS}{orthogonal time frequency space}
    \acro{i.i.d.}{independent and identically distributed}
    \acro{LTE}{long term evolution}
    \acro{ISFFT}{inverse symplectic fast Fourier transform}
    \acro{SFFT}{symplectic fast Fourier transform}
    \acro{B5G}{beyond-5G}
    \acro{FFT}{fast Fourier transform}
    \acro{IFFT}{inverse fast Fourier transform}
    \acro{DFT}{discrete Fourier transform}
    \acro{OMP}{orthogonal matching pursuit}
    \acro{AMP}{approximate message passing}
    \acro{LMMSE}{linear minimum mean square error}
    \acro{CSI}{channel state information}
    \acro{PN}{pseudo-noise}
    \acro{MUI}{multi-user interference}
    \acro{MA}{multiple access}
    \acro{DD}{delay-Doppler}
    \acro{TF}{time-frequency}
    \acro{NMSE}{normalized mean square error}
    \acro{CS}{compressive sensing}
    \acro{BPSK}{binary phase-shift keying}
    \acro{QPSK}{quadrature phase-shift keying}
    \acro{SCI}{side channel information}
    \acro{LS}{least-squares}
    \acro{BER}{bit error rate}
    \acro{IDZT}{inverse discrete Zak transform}
\end{acronym}

\begin{abstract}
Channel estimation techniques for orthogonal time frequency space (OTFS) modulation scheme are investigated. The orthogonal matching pursuit algorithm is investigated with and without side channel information, and an efficient data placement is proposed alongside the pilot in the multi-user scenario based on impulse pilot-based estimation. Finally, the performance of the estimation techniques across different multi-user scenarios is evaluated and compared, highlighting the strengths and weaknesses of each method.
\end{abstract}

\begin{IEEEkeywords}
Orthogonal Time Frequency Space Modulation, Orthogonal Matching Pursuit, Channel Estimation
\end{IEEEkeywords}

\section{Introduction}
\label{sec:intro}
Massive \ac{MIMO} technology combines high data rates with efficient processing \cite{biglieri2007mimo}. For a large geographic region, one generally uses distributed antenna structures, called \acp{AP}, connected by backhaul links \cite{ngo2015cell}. The distributed structure provides macro-diversity gain by allowing a large number of \acp{AP} to communicate with a relatively smaller number of \acp{UE}, unlike traditional massive MIMO where each \ac{UE} is served by a single base station. The system has no cell boundaries and is called cell-free massive \ac{MIMO} \cite{demir2021foundations}. The \acp{AP} are often assumed to be single antenna devices that serve users in a \ac{TDD} mode. Future networks may need advanced modulation schemes to support time and frequency-varying channels, also called doubly-selective channels. \ac{OTFS} modulation\cite{hadani2018otfs,hadani2017orthogonal} can provide higher data rates than \ac{OFDM} in scenarios with high Doppler shifts \cite{lopez2022achievable, an2019high, raviteja2019otfs, mohammadi2022performance, gaudio2021otfs}. \ac{OTFS} can be considered as a precoded \ac{OFDM} where the information symbols are placed in the \ac{DD} domain. The information symbols are transformed to the \ac{TF} domain through the Zak transform and are sent to the receiver with \ac{OFDM} modulation \cite{lampel2022otfs, 9927420}.

One challenge in cell-free massive \ac{MIMO} systems is accurate channel estimation \cite{Guo_2022,song2021joint}. As users are served by multiple distributed \acp{AP}, the \ac{CSI} becomes complex due to more diverse interference and channel characteristics. The paper \cite{murali2018otfs} estimates the delay, Doppler, and channel gains using \ac{PN} sequences as pilots. This method has large complexity because it estimates the delay-Doppler coefficients in the \ac{TF} domain. 
The paper \cite{ramachandran2018mimo} has each antenna transmit a single pilot impulse in a frame to estimate the channel. This leads to a large loss in spectral efficiency because most frames containing the pilot cannot carry data. Moreover, this method may generate errors in estimation when applied to rapidly time-varying channels because the channel information estimated by the pilot in each frame is utilized for data detection in the subsequent frame. 
Accordingly, \cite{surabhi2019multiple} places multiple pilot impulses in a single frame with sufficient guard bands depending on the maximum delay and Doppler spreads associated with the channel. 

An important aspect of \ac{OTFS} modulation is that it has a sparse \ac{DD} channel representation \cite{hadani2017orthogonal}. This sparsity is utilized by a variety of algorithms. For example, the papers \cite{wei2022off, srivastava2021bayesian} propose a sparse Bayesian learning method, \cite{li2019low} improved the complexity of \ac{AMP}, and \cite{rasheed2020sparse} uses \ac{OMP} for multiple users. To further improve the spectral efficiency, the paper \cite{raviteja2019embedded} introduced an embedded pilot-aided channel estimation, where the data and pilot symbols are placed in the same frame. Similarly, \cite{mishra2021otfs} placed both data and pilot symbols in the same frame with different power levels.


The main contributions of this paper are as follows:
\begin{itemize}
    \item a comparison of \ac{OMP} and threshold-based channel estimation for different multi-user scenarios;
    \item a study of \ac{OMP}-based channel estimation with \ac{SCI};
    \item a bandwidth-efficient pilot and data placement for multi-user channel estimation and data detection.
\end{itemize}

\subsubsection*{Notation} We use the following general notations. $(\cdot)^\ast$, $(\cdot)^\transp$, and $(\cdot)^\herm$ denote the complex conjugate, transpose, and Hermitian (conjugate and transpose) operations, respectively. $\left|x\right|$ is the absolute value of $x\in\RR$ and $|\Xc|$ denotes the cardinality of the set $\Xc$. $\|\xv\|_p$ is the $\ell_p$-norm of a complex or real vector $\xv$. $\Fm_{\rm D}$ is the $\rm D$-dimensional \ac{DFT} matrix while $\Gm_{\rm tx/rx}$ is the diagonal matrix containing the pulse-waveform samples at the \ac{Tx} and \ac{Rx}, respectively. The $m \times m$ identity matrix is written as $\Id_m$. The variables $M$ and $N$ represent the number of bins along the delay axes and the Doppler axes, respectively.

\section{System Model}
\label{sec:sys}

Consider an \ac{AP} and \acp{UE} $u \in \{1, \dots, U\}$ that communicate over a sparse channel in DD domain with $L$ distinct paths as follows
\begin{align}\label{eq:multi_channel_cont_domain}
    h_u(\tau, \nu) = \sum_{i = 1}^{L} h_{i,u} \delta(\tau - \tau_{i,u}) \delta(\nu - \nu_{i,u}),
\end{align}
where $h_{i,u}$, $\tau_{i,u}$, and $\nu_{i,u}$ are the channel coefficient, delay, and Doppler shift associated with the $i$-th path and the user $u$, respectively.

In the OTFS modulation, the symbols are placed in the DD domain. Assume the total duration $NT$ and bandwidth $M\Delta f$ for the transmitted signal frame such that the pulse shaping waveforms are sampled at the Nyquist rate $T/M$ where $T=1/\Delta f$. Then, similar to \cite{raviteja2018practical}, we assume that the delay and Doppler values are integer shifts of $1/(M \Delta f)$ and $1/(NT)$ respectively, as follows 
\begin{align}
    \tau_{i,u} = \frac{l_{i,u}}{M\Delta f} =\frac{T}{M}l_{i,u}, \quad \nu_{i,u} = \frac{k_{i,u}}{NT} = \frac{\Delta f}{N}k_{i,u},\label{eq:dd_tap}
\end{align}
where $l_{i,u}\in \{0, \dots, l_{\max}-1\}$ and $k_{i,u}\in \{0, \dots, k_{\max}-1\}$ are integer values corresponding to the delay and Doppler for path $i$ and user $u$, respectively. Moreover, we assume $l_{\max} \le M$ and $k_{\max} \le N$ are the maximum shift in delay and Doppler in all paths and for all users.  Thus, the symbols are placed in a $M\times N$ grid $\Xm_{\mathrm{DD},u}(m,n)$, $m\in \{0,\dots, M-1\}$, $n \in \{0, \dots, N-1\}$, in the DD domain with the delay $1/(M\Delta f)$ and Doppler $1/(NT)$ resolutions\footnote{In this work, we do not assume fractional delay and Doppler, see \cite{raviteja2018interference, fish2013delay}.}, as shown in Fig. \ref{fig:impulse_pilot_entire_grid}.




Let $s_u(t)$ be the transmitted signal and $r_u(t)$ denote the received signal after sampling and discarding \ac{CP} from the user $u$. Moreover, assume the superposition of the users is $r(t)=\sum_{u=1}^U r_u(t)$. Then, we have
\begin{align} \label{eq:rx_multi_cont}
     r_u(t)&= \int_{\nu} \int_{\tau} h_u(\tau, \nu) s_u(t - \tau) e^{j 2 \pi \nu (t - \tau)} d\tau d\nu\nonumber\\
     &\quad+ w_u(t),
\end{align}
where $w_u(t)$ is the white Gaussian noise with power spectral density $N_0$.
Now, by sampling at the rate $M\Delta f=M/T$ and substituting \eqref{eq:multi_channel_cont_domain} and \eqref{eq:dd_tap} into \eqref{eq:rx_multi_cont}, the discrete received signal is
\begin{align}
     r_u(n)&= \sum_{i = 1}^{L} h_{i,u} s([n - l_{i,u}]_{MN}) e^{j 2 \pi \frac{k_{i,u}}{MN}(n - l_{i,u})} \nonumber\\
     &\quad+ w_u(n),
\end{align}
where $[\cdot]_n$ is the mod-$n$ operation, see \cite{raviteja2018practical}.
Thus, the sampled received signal in the discrete time domain is
\begin{align} \label{eq:sampled_received_signal_TF}
    \rv_u = \Hm_u \sv_u + \wv_u,
\end{align}
where $\sv_u^{MN\times 1}$ is the discrete transmit vector, $\wv_u$ is the noise vector, and $\Hm_u^{MN\times MN}$ is the time-domain channel matrix as follows 
\begin{align}\label{eq:multi_channel_disc_domain}
    \Hm_u = \sum_{i = 1}^{L} h_{i,u} \Pim^{l_{i,u}} \Deltam^{k_{i,u}},
\end{align}
where $\Pim \in \RR^{MN \times MN}$ is the permutation matrix
\begin{align}
\Pim = \begin{bmatrix}
0 & 0 & 0 & \dots & 1 \\
1 & 0 & 0 & \dots & 0 \\
0 & 1 & 0 & \dots & 0 \\
\vdots & \vdots & \vdots & \ddots & \vdots \\
0 & 0 & \dots & 1 & 0
\end{bmatrix}
\end{align}
and $\Deltam \in \CC^{MN \times MN}$ is a diagonal matrix with the i-th entry $\Delta_i=z^{i}$, $i \in [MN]$, and $z=\exp{(j2\pi/MN)}$, see \cite{raviteja2018practical}.

Finally, assume the diagonal matrices 
\begin{align}
    \Gm_{\rm{tx}}&=\diag\left(g_{tx}(0), g_{tx}\left({T/M}\right), \dots, g_{tx}\left({(M-1)T/M}\right)\right),\nonumber\\
    \Gm_{\rm{rx}}&=\diag\left(g_{rx}(0), g_{rx}\left({T/M}\right), \dots, g_{rx}\left({(M-1)T/M}\right)\right),\nonumber
\end{align}
where $g_{tx}(.)$ and $g_{rx}(.)$ are the transmit pulse shape and the filter at the receiver. Then, equation \eqref{eq:sampled_received_signal_TF} in the DD domain is as follows
\begin{align} \label{eq:end_to_end_output}
    \yv_{\mathrm{DD}, u} = \Hm_{\mathrm{eff}, u} \xv_{\mathrm{DD}, u} + \wv_{\mathrm{DD, u}},
\end{align}
where $\yv_{\mathrm{DD}, u}=\left(\Fm_{\rm{N}} \otimes \Gm_{rx}\right)\rv_u$, $\xv_{\mathrm{DD}, u}=\Vec{\left(\Xm_{\mathrm{DD}, u}\right)}$ is the column-wise vectorization of DD domain gird, $\wv_{\mathrm{DD}, u} = \left(\Fm_{\rm{N}} \otimes \Gm_{rx}\right) \wv_{u}$ is the effective noise vector in the DD domain, $\Hm_{\mathrm{eff}, u} = \left(\Fm_{\rm{N}} \otimes \Gm_{rx}\right)\Hm_u \left(\Fm_{\rm{N}}^{\herm} \otimes \Gm_{tx}\right)$ is the effective end-to-end \ac{DD} domain channel matrix corresponding to user $u$, $\Fm_{\rm N}$ is the $\rm N$-point FFT, and $\otimes$ is the Hadamard product, see \cite{raviteja2018practical}. 
Moreover, assuming $\yv_{\mathrm{DD}, u}=\Vec{\left(\Ym_{\mathrm{DD}, u}\right)}$ and $\wv_{\mathrm{DD}, u}=\Vec{\left(\Wm_{\mathrm{DD},u}\right)}$, equation \eqref{eq:end_to_end_output} can be expressed with a 2D convolution operation. In particular, $\Ym_{\mathrm{DD}, u}(m,n)$ is
\begin{align} \label{eq:rx_signal_DD}
    &\sum_{i=1}^{L}h_{i,u} \alpha_{i,u}(m,n) \cdot \Xm_{\mathrm{DD},u} ([m-l_{i,u}]_{M},[n-k_{i,u}]_{N})\nonumber\\
    &\quad+ \Wm_\mathrm{DD}(m,n),
\end{align}
where 
\begin{align}
    \alpha_{i,u}(m,n) = 
    \begin{cases}
        e^{-j2\pi\frac{n}{N}}z^{k_{i,u}([m-l_{i,u}]_{M})} & \text{if } m < l_{i,u} \\
        z^{k_{i,u}([m-l_{i,u}]_{M})} & \text{if } m \geq  l_{i,u}\\
        0 & \text{else}.
    \end{cases}
\end{align}
is the correction factor which accounts for the phase deviations caused by the loss of bi-orthogonality between the \ac{Tx} and \ac{Rx} pulses; see \cite{raviteja2018practical}.

\section{OTFS Channel Estimation}
\label{sec:otfs_channel_estimate}
\subsection{Compressive Sensing based Channel Estimation}
By exploiting sparsity, \ac{CS} permits channel parameter estimation with reduced sampling requirements.
We use the \ac{OMP} algorithm due to its reconstruction accuracy and low computational complexity \cite{draganic2017some}. We compare the obtained estimation error with that of the modified \ac{OMP} algorithm that considers \ac{SCI} for the purpose of estimation. 
To reduce the complexity, we consider the placement of pilot symbols in the time domain by allocating the $MN$ resource vector across multiple users in the uplink.

Combine \eqref{eq:sampled_received_signal_TF} and \eqref{eq:multi_channel_disc_domain} to obtain
\begin{align}\label{eq:received_signal-01}
    \rv_u = \left(\sum_{i = 1}^{L} h_{i,u} \Pim^{l_{i,u}} \Deltam^{k_{i,u}}\right) \sv_u + \wv_u.
\end{align}
Equivalently, we can convert the summation in \eqref{eq:received_signal-01} to the values of delay and Doppler in each path as follows
\begin{align}\label{eq:received_signal-02}
    \rv_u = \sum_{l_u = 0}^{l_{\max}-1} \sum_{k_u = 0}^{k_{\max}-1}  h_{l_u, k_u} \Pim^{l_{u}} \Deltam^{k_{u}} \sv_u + \wv_u.
\end{align}
Finally, defining $\Psim_{l_{u},k_{u}}:=\Pim^{l_{u}} \Deltam^{{k_{u}}} \sv_{u}$, \eqref{eq:received_signal-02} can be written as
\begin{align} \label{eq:received_signal}
    &\rv_u = \Psim_u \hv_u + \wv_u,\quad u\in [U],
\end{align}
where
\begin{align}
\Psim_u &= \left[\Psim_{0,0}  \Psim_{1,0}  \dots  \Psim_{M-1,0}  \Psim_{0,1}  \dots  \Psim_{M-1,N-1}\right],\nonumber\\
\hv_u^{\transp} &= \left[h_{0,0}  h_{1,0}  \dots  h_{M-1,0}  h_{0,1}  \dots  h_{M-1,N-1}\right].
\end{align}
The vector $\hv_u^{MN \times 1}$ is the sparse channel coefficient and $\Psim_u$ is the sensing matrix associated with the $u$-th user. Thus, the \ac{OTFS} uplink channel estimation is formulated as a sparse recovery problem:
\begin{align}
    \min \|\hv\|_{0} \quad
    \textrm{s.t.} \quad  \rv = \Psim\hv + \wv,\label{eq:objective-sparse}
\end{align}
where $\rv=\sum_{u=1}^U \rv_u$, $\wv=\sum_{u=1}^U \wv_u$, and
\begin{align}
\Psim = \left[\Psim_1  \Psim_2  \dots  \Psim_U \right],\quad \hv^T = \left[\hv_1  \hv_2  \dots  \hv_U \right].
\end{align}

Finally, Algorithm \ref{OMP_algo} solves \eqref{eq:objective-sparse} with the \ac{OMP} \cite{rasheed2020sparse}. In this algorithm, the parameters $\epsilon$, $\Sm^n$, $\qv$, and $T^n$ are the sparsity threshold, the selected support set, the residual vector and the index corresponding to the maximum inner product between the current residual and the adjoint of the sensing matrix, respectively. Furthermore, if $L$ is known, Algorithm \ref{OMP_algo} can be modified to Algorithm \ref{OMP_SCI_algo} which has a better computational complexity independent of the threshold; see Sec. \ref{sec:numerical}.


\begin{algorithm}[t]
    \SetAlgoLined\usetikzlibrary{shapes.geometric}
\usetikzlibrary{arrows.meta}
    \DontPrintSemicolon
    \SetKw{KwInit}{Initialize:}
     \KwIn{$\rv, \Psim$,$\epsilon$}
     \KwOut{$\Sm^n, \hv_{\Sm^n}$}
     \KwInit{$n = 0$, $\hv^{0}= \mathbf{0}$, $\Sm^{0}$ = $\emptyset$, $\qv^{0}= \rv$}\\
     \While{($|\norm\qv_{n-1}^{2}-\norm\qv_{n}^{2}| > \epsilon$)}{
     $n = n + 1$ \\
     $T^{n} =   \argmax \lvert{\Psim^{\herm} \qv^{n-1}}\rvert$ \\
     $\Sm^{n} = \Sm^{n-1} \cup T^{n}$\\
     $\hv_{S^{n}} = {\Psim^{\dagger}_{\Sm^{n}}\rv}$\\
     $\qv^{n-1} = \qv^{n}$\\
     $\qv^{n} = \rv - \Psim_{\Sm^{n}}\hv_{\Sm^{n}}$
     }
     \caption{\ac{OMP} Algorithm.}
     \label{OMP_algo}
\end{algorithm}

\begin{algorithm}[t]
    \SetAlgoLined
    \DontPrintSemicolon
    \SetKw{KwInit}{Initialize:}
     \KwIn{$\rv, \Psim$, $c=L$}
     \KwOut{$\Sm^n, \hv_{\Sm^n}$}
     \KwInit{$n = 0$, $\hv^{0}= \mathbf{0}$, $\Sm^{0}= \emptyset$, $\qv^{0} = \rv$}\\
     \While{($c > 0$)}{
     $n = n + 1$ \\
     $T^{n} =   \argmax \lvert{\Psim^{\herm}\qv^{n-1}}\rvert$ \\
     $\Sm^{n} = \Sm^{n-1} \cup T^{n}$\\
     $\hv_{S^{n}} = {\Psim^{\dagger}_{\Sm^{n}}\rv}$\\
     $\qv^{n} = \rv - \Psim_{\Sm^{n}}\hv_{\Sm^{n}}$\\
     $c = c-1$
     }
     \caption{\ac{OMP} with \ac{SCI}.}
     \label{OMP_SCI_algo}
\end{algorithm}
\subsection{Impulse Pilot-based Channel Estimation}
\label{sec:impulse-pilot}
\begin{figure}[t]
    \centering
    \includegraphics[width = 0.4\textwidth]{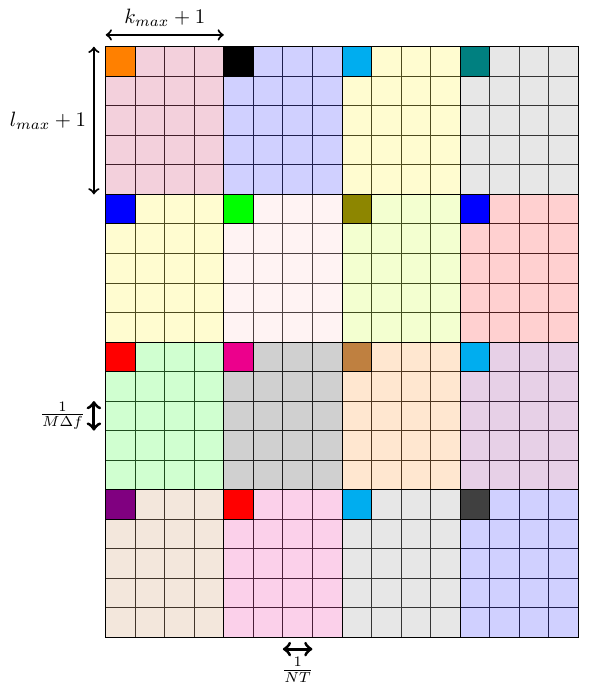}
    \caption{DD grid with maximum possible pilot symbols with associated guard bands.}
    \label{fig:impulse_pilot_entire_grid}
\end{figure}

Let $l_{\max}$ and $k_{\max}$ be the maximum delay and Doppler in each path $i$ and for each user $u$. Each user $u$ places a pilot $p_u$ in the \ac{DD} domain $\mathbf{X}_{\mathrm{DD},u}(m,n)$ with sufficient guard band $l_{\max}$ and $k_{\max}$ as shown in Fig. \ref{fig:impulse_pilot_entire_grid}. Thus, the channel estimation can be performed without multi-user interference on the received grid $\mathbf{Y}_\mathrm{DD}(m,n)$, see \eqref{eq:rx_signal_DD}. In particular, the threshold-based channel estimation $\hat{h}_{i,u}$ and the detection of delay $l_{i,u}$ and Doppler $k_{i,u}$ associated with the $i$-th path and user $u$ at location $(l_{i,u}, k_{i,u})$ is
\begin{align}
    \hat{h}_{i,u} = \frac{\Ym_{\rm DD}(l_{i,u}, k_{i,u})}{p_{u}  \alpha_{i,u}(l_{i,u},k_{i,u})} \quad \text{if} \quad \Ym_{\rm DD}(l_{i,u}, k_{i,u}) > \tau,
\end{align}
for some threshold $\tau$.
The overall number of symbols for pilot and guard band needed to have a perfect channel estimation at the \ac{AP} is $(l_{\max}+1)(k_{\max}+1)$ per user. Thus, for an $M \times N$ grid, the maximum number of users that can be estimated simultaneously and independently is $\lfloor MN/(l_{\max}+1)(k_{\max}+1) \rfloor$; see Fig.~(\ref{fig:impulse_pilot_entire_grid}).
\subsection{Embedded Data Transmission and Channel Estimation}

The estimation method in Sec.~\ref{sec:impulse-pilot} uses the entire grid for channel estimation, which generally reduces spectral efficiency. Instead, when serving a small number of users, one may fill the empty \ac{DD} grid with information symbols to improve spectral efficiency. We propose a strategic data placement alongside the pilots to increase spectral efficiency. 

\begin{figure}[t]
    \centering
    \scalebox{0.7}{\begin{tikzpicture}
    
    \foreach \x in {0,0.5,1,1.5,2,2.5,3,3.5,4,4.5,5,5.5,6,6.5,7,7.5}
        \foreach \y in {-0.5,0,0.5,1,1.5,2,2.5,3,3.5,4,4.5,5,5.5,6,6.5,7,7.5,8,8.5,9}
            \filldraw[fill=blue!50, draw=black, fill opacity=0.3] (\x,\y) rectangle (\x+0.5,\y+0.5);
            
    \filldraw[fill=purple!100, draw=black, fill opacity=0.3] (0,9.5) rectangle (2,7);
    \filldraw[fill=orange!100, draw=black] (0,9) rectangle (0.5,9.5);

    \filldraw[fill=purple!100, draw=black, fill opacity=0.3] (6.5,9.5) rectangle (8,7);
    
    
    
    \filldraw[fill=purple!100, draw=black, fill opacity=0.3] (0,1.5) rectangle (2,-0.5);



    \filldraw[fill=purple!100, draw=black, fill opacity=0.3] (6.5,1.5) rectangle (8,-0.5);
    
    \draw[<->,line width=1pt] (0,9.7) -- (2,9.7) node[midway, above] {$k_{max}+1$};
    \draw[<->,line width=1pt] (-0.2,9.5) -- (-0.2,7) node[midway, left] {$l_{max}+1$};

    \draw[<->,line width=1pt] (0,-0.7) -- (2,-0.7) node[midway, below] {$k_{max}+1$};
    \draw[<->,line width=1pt] (-0.2,1.5) -- (-0.2,-0.5) node[midway, left] {$l_{max}$};

    \draw[<->,line width=1pt] (6.5,-0.7) -- (8,-0.7) node[midway, below] {$k_{max}$};
    \draw[<->,line width=1pt] (8.2,1.5) -- (8.2,-0.5) node[midway, right] {$l_{max}$};

    \draw[<->,line width=1pt] (6.5,9.7) -- (8,9.7) node[midway, above] {$k_{max}$};
    \draw[<->,line width=1pt] (8.2,9.5) -- (8.2,7) node[midway, right] {$l_{max}+1$};

    \draw[<->,line width=1.5pt] (3,-0.7) -- (3.5,-0.7) node[midway, below] {$\frac{1}{NT}$};
    \draw[<->,line width=1.5pt] (-0.2,3) -- (-0.2,3.5) node[midway, left] {$\frac{1}{M\Delta f}$};

\end{tikzpicture}}
    \caption{DD grid with pilot and data for a single user.}
    \label{fig:embedded_pilot_user_1}
\end{figure}
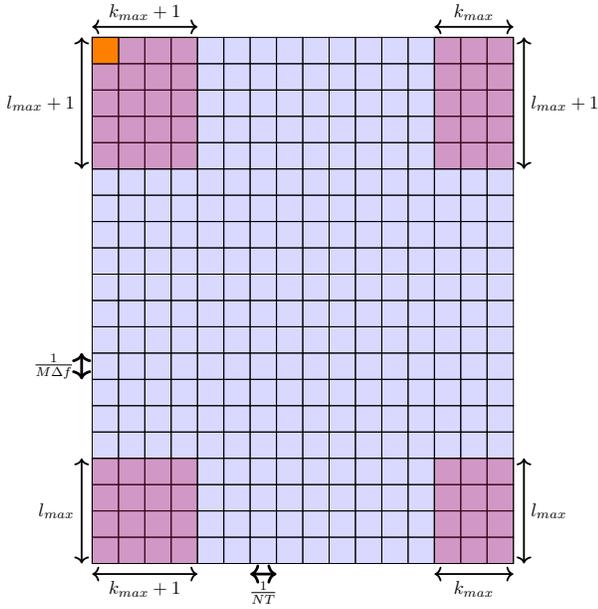

\begin{figure}[t]
    \centering
    \scalebox{0.7}{\begin{tikzpicture}
    
    \foreach \x in {0,0.5,1,1.5,2,2.5,3,3.5,4,4.5,5,5.5,6,6.5,7,7.5}
        \foreach \y in {-0.5,0,0.5,1,1.5,2,2.5,3,3.5,4,4.5,5,5.5,6,6.5,7,7.5,8,8.5,9}
            \filldraw[fill=blue!50, draw=black, fill opacity=0.3] (\x,\y) rectangle (\x+0.5,\y+0.5);
            
    \filldraw[fill=purple!100, draw=black, fill opacity=0.3] (0,9.5) rectangle (2,7);
    \filldraw[fill=orange!100, draw=black] (0,9) rectangle (0.5,9.5);
            
    \filldraw[fill=black!60, draw=black, fill opacity=0.3] (0,7) rectangle (2,4.5);
    \filldraw[fill=blue!100, draw=black] (0,6.5) rectangle (0.5,7);

    \draw[<->,line width=1pt] (-0.2,7) -- (-0.2,4.5) node[midway, left] {$l_{max}+1$};


    \filldraw[fill=yellow!60, draw=black, fill opacity=0.3] (2,9.5) rectangle (4,7);
    \filldraw[fill=black!100, draw=black] (2,9) rectangle (2.5,9.5);

    \draw[<->,line width=1pt] (2,9.7) -- (4,9.7) node[midway, above] {$k_{max}+1$};

    \filldraw[fill=orange!60, draw=black, fill opacity=0.3] (2,7) rectangle (4,4.5);
    \filldraw[fill=green!100, draw=black] (2,6.5) rectangle (2.5,7);





    \filldraw[fill=purple!100, draw=black, fill opacity=0.3] (6.5,9.5) rectangle (8,7);
    
    \filldraw[fill=black!60, draw=black, fill opacity=0.3] (6.5,7) rectangle (8,4.5);
    
    
    \filldraw[fill=purple!100, draw=black, fill opacity=0.3] (0,1.5) rectangle (2,-0.5);

    \filldraw[fill=yellow!60, draw=black, fill opacity=0.3] (2,1.5) rectangle (4,-0.5);


    \filldraw[fill=purple!100, draw=black, fill opacity=0.3] (6.5,1.5) rectangle (8,-0.5);
    
    \draw[<->,line width=1pt] (0,9.7) -- (2,9.7) node[midway, above] {$k_{max}+1$};
    \draw[<->,line width=1pt] (-0.2,9.5) -- (-0.2,7) node[midway, left] {$l_{max}+1$};

    \draw[<->,line width=1pt] (0,-0.7) -- (2,-0.7) node[midway, below] {$k_{max}+1$};
    \draw[<->,line width=1pt] (-0.2,1.5) -- (-0.2,-0.5) node[midway, left] {$l_{max}$};

    \draw[<->,line width=1pt] (6.5,-0.7) -- (8,-0.7) node[midway, below] {$k_{max}$};
    \draw[<->,line width=1pt] (8.2,1.5) -- (8.2,-0.5) node[midway, right] {$l_{max}$};

    \draw[<->,line width=1pt] (6.5,9.7) -- (8,9.7) node[midway, above] {$k_{max}$};
    \draw[<->,line width=1pt] (8.2,9.5) -- (8.2,7) node[midway, right] {$l_{max}+1$};

    \draw[<->,line width=1.5pt] (3,-0.7) -- (3.5,-0.7) node[midway, below] {$\frac{1}{NT}$};
    \draw[<->,line width=1.5pt] (-0.2,3) -- (-0.2,3.5) node[midway, left] {$\frac{1}{M\Delta f}$};

\end{tikzpicture}}
    \caption{DD grid with pilot and data for four users.}
    \label{fig:embedded_pilot_four_user}
\end{figure}
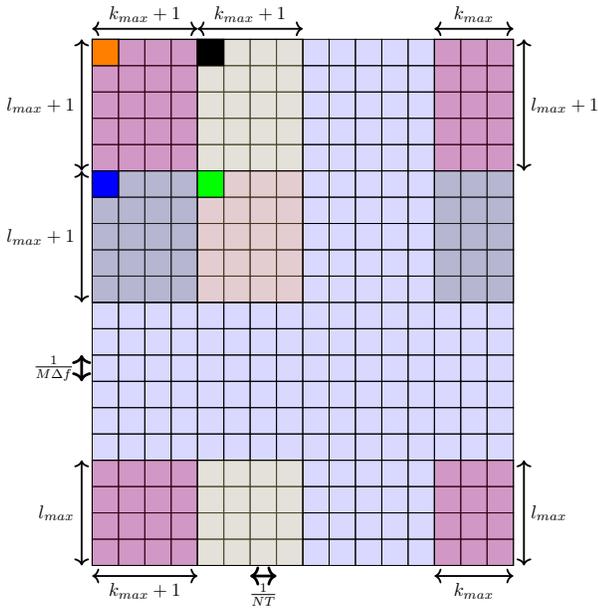

Consider a pilot symbol located at (0,0) of the \ac{DD} grid and assume $l_{\max}$ and $k_{\max}$ are the maximum delay and Doppler shifts; see Fig.~\ref{fig:embedded_pilot_user_1}. We need a guard band at all corners to avoid data collisions, in addition to the guard band of size ($l_{\max}+1$)($k_{\max}+1$) in the upper left corner. This gives a total guard band of size $4 l_{\max} k_{\max} + 2 l_{\max} + 2 k_{\max} + 1$.
By adding a second user's pilot below the first user's guard band, see Fig.~\ref{fig:embedded_pilot_four_user}, we need an additional guard band with $2 l_{\max} k_{\max} + l_{\max} + 2 k_{\max} + 1$ symbols. The reduction is due to the presence of the guard band associated with the first user. Similarly, by adding a third user, we need an additional guard band with $2 l_{\max} k_{\max} + 2 l_{\max} + k_{\max} + 1$ symbols. Finally, the fourth user's pilot can be added with an additional guard band of $l_{\max} k_{\max} + l_{\max} + k_{\max} + 1$ symbols.  This process can be continued until all user's pilots are placed; see Fig. \ref{fig:embedded_pilot_four_user}.


The \ac{AP} receives the superposition of all users' transmitted signals as follows
\begin{align} \label{eq:data_detection_received_vector}
    \yv_{\rm DD} = \Hm_{\rm eff} \xv_{\rm DD} + \wv_{\rm DD}
\end{align}
where $\yv_{\rm DD}=\sum_{u=1}^U \yv_{\rm{DD}, u}$, $\wv_{\rm DD}=\sum_{u=1}^U \wv_{\rm{DD}, u}$, 
\begin{align}
    \Hm_{\rm eff} &= [\Hm_{\rm eff, 1}, \dots, \Hm_{\rm eff, U}], && \Hm_{{\rm eff}, u} \in \CC^{MN \times MN},\\
    \xv^\transp_{\rm DD} & = [\xv_{\rm DD,1}, \dots, \xv_{\rm DD, U}], && \xv_{{\rm DD}, u} \in \CC^{MN \times 1}.
\end{align}
After the channel estimation, the \ac{AP} will perform multi-user detection by extending the linear estimation technique mentioned in \cite{mehrotra2022channel}. The sub-matrices corresponding to data indices for the estimated channel matrix $\hat{\Hm}_{\rm eff}$ can be computed as
\begin{align}
    \yv_{\rm DD,\mathcal{D}} = \hat{\Hm}_{\rm eff,\mathcal{D}} \xv_\mathrm{DD,\mathcal{D}} + \wv_{\rm{DD}, \mathcal{D}},
\end{align}
where $\xv_\mathrm{DD,\mathcal{D}}$ and $\wv_{\rm{DD}, \mathcal{D}}$ are obtained by removing the data in the indices corresponding to the location of pilots, and $\hat{\Hm}^{\prime}_{\rm eff,\mathcal{D}}$ is obtained by removing the corresponding columns. Finally, the detection can be accomplished by an \ac{LMMSE} estimator as follows
\begin{align}
    \hat{\xv}^{\rm MMSE}_{\rm DD,\mathcal{D}} = (\hat{\Hm}^{\herm}_{\mathrm{eff},\mathcal{D}} \hat{\Hm}_{\mathrm{eff},\mathcal{D}} + \frac{1}{\rm SNR} \Id)^{-1} \hat{\Hm}^{\herm}_{\rm eff, \mathcal{D}} \yv_{\rm DD,\mathcal{D}}.\\\nonumber
\end{align}


\section{Numerical Results}
\label{sec:numerical}
The simulation is done with a $(M \times N) = (32 \times 32)$ \ac{DD} grid and the carrier frequency $f_c = \SI{4}{\giga\hertz}$ with a sub-carrier spacing of $\Delta f = \SI{15}{\kilo\hertz}$. Moreover, we assume an additive circularly symmetric Gaussian noise with variance $\sigma=1$, \ac{BPSK} modulation for the pilot symbols, and we consider $L=4$ \ac{i.i.d.} paths with a random power-delay profile. This model includes scatterers in the communication environment, adding a layer of realism. We investigate the robustness of the above \ac{CS}-based and impulse-based channel estimation algorithms.

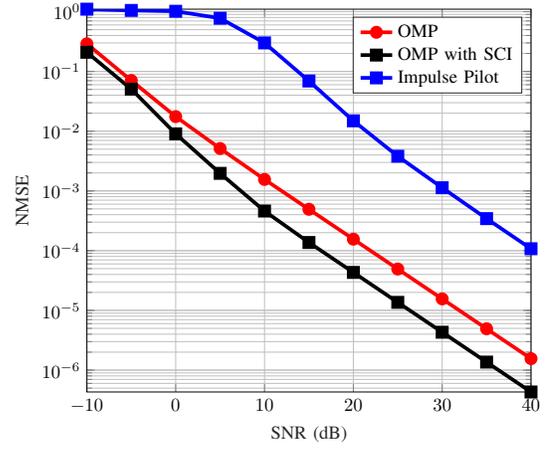
\begin{figure}[t]
    \centering
    \scalebox{0.8}{
%
%
\definecolor{mycolor1}{rgb}{1.00000,0.00000,1.00000}%
\begin{tikzpicture}

\begin{axis}[%
scale only axis,
xmin=-10,
xmax=40,
xlabel style={font=\color{white!15!black}},
xlabel={SNR (dB)},
ymode=log,
ymin=4.30980708863487e-07,
ymax=1.10134880609495,
yminorticks=true,
ylabel style={font=\color{white!15!black}},
ylabel={NMSE},
axis background/.style={fill=white},
title style={font=\bfseries},
xmajorgrids,
ymajorgrids,
yminorgrids,
grid style={line width=0.3pt, draw=gray!20},
legend style={legend cell align=left, align=left, draw=white!15!black}
]
\addplot [color=red, line width=2.0pt, mark size=1.4pt, mark=*, mark options={solid, fill=red, red}]
  table[row sep=crcr]{%
-10	0.290170017900149\\
-5	0.0704837643624645\\
0	0.0175302976017404\\
5	0.00512841867597555\\
10	0.00155729948554356\\
15	0.000492461337332611\\
20	0.000155729948554357\\
25	4.92461337332612e-05\\
30	1.55729948554358e-05\\
35	4.92461337332601e-06\\
40	1.55729948554356e-06\\
};
\addlegendentry{OMP}



\addplot [color=black, line width=2.0pt, mark size=1.4pt, mark=square*, mark options={solid, fill=black, black}]
  table[row sep=crcr]{%
-10	0.209334609210166\\
-5	0.0505436225962646\\
0	0.00902150235761862\\
5	0.00196854580555567\\
10	0.000460687347884076\\
15	0.000136988373575218\\
20	4.3226156664624e-05\\
25	1.36288066760252e-05\\
30	4.30980708863454e-06\\
35	1.36288066760248e-06\\
40	4.30980708863487e-07\\
};
\addlegendentry{OMP with SCI}

\addplot [color=blue, line width=2.0pt, mark size=1.4pt, mark=square*, mark options={solid, fill=blue, blue}]
  table[row sep=crcr]{%
-10	1.10134880609495\\
-5	1.04989395887777\\
0	1.01659201715646\\
5	0.776778136164128\\
10	0.300421416864257\\
15	0.0691906853495896\\
20	0.0149032145641843\\
25	0.00380629142010903\\
30	0.00112961828573473\\
35	0.000343766631679001\\
40	0.000107684200934039\\
};
\addlegendentry{Impulse Pilot}

\end{axis}

\end{tikzpicture}%
}
    \caption{\ac{NMSE} comparison for four users.}
    \label{fig:CER_4_Users}
\end{figure}

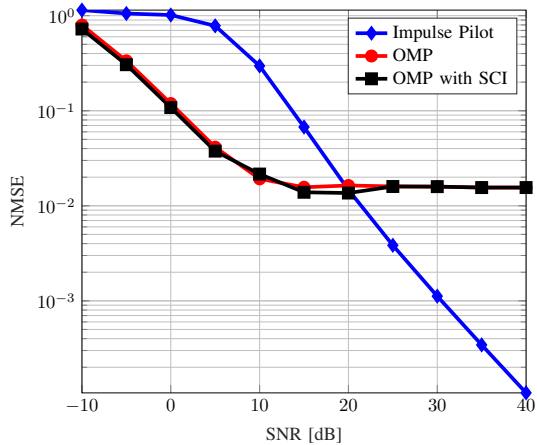
\begin{figure}[t]
    \centering
    \scalebox{0.8}{
%
%

\begin{tikzpicture}

\begin{axis}[%
scale only axis,
xmin=-10,
xmax=40,
xlabel style={font=\color{white!15!black}},
xlabel={SNR [dB]},
ymode=log,
ymin=0.000107539804522811,
ymax=1.14508018749548,
yminorticks=true,
ylabel style={font=\color{white!15!black}},
ylabel={NMSE},
axis background/.style={fill=white},
xmajorgrids,
ymajorgrids,
yminorgrids,
grid style={line width=0.3pt, draw=gray!20},
legend style={legend cell align=left, align=left, draw=white!15!black}
]
\addplot [color=blue, line width=2.0pt, mark size=1.4pt, mark=diamond*, mark options={solid, fill=blue, blue}]
  table[row sep=crcr]{%
-10	1.14508018749548\\
-5	1.05448985245101\\
0	1.01696732531535\\
5	0.782212150165738\\
10	0.296158909354007\\
15	0.0673519538701832\\
20	0.0147277529086636\\
25	0.0038389680208769\\
30	0.00111947113602578\\
35	0.000343873469174326\\
40	0.000107539804522811\\
};
\addlegendentry{Impulse Pilot}

\addplot [color=red, line width=2pt, mark size=1.4pt, mark=*, mark options={solid, fill=red, red}]
  table[row sep=crcr]{%
-10	0.796822004343958\\
-5	0.335231981570585\\
0	0.119138349591419\\
5	0.041361369309223\\
10	0.0192079161850394\\
15	0.0156658883698194\\
20	0.0163582416079339\\
25	0.0159992481974969\\
30	0.015929400450926\\
35	0.0155560269530312\\
40	0.0155608280651173\\
};
\addlegendentry{OMP}

\addplot [color=black, line width=2pt, mark size=1.4pt, mark=square*, mark options={solid, fill=black, black}]
  table[row sep=crcr]{%
-10	0.723932219702112\\
-5	0.305083234589632\\
0	0.108057877836564\\
5	0.0375921816156802\\
10	0.0215734992268564\\
15	0.0138736498919961\\
20	0.0136224087438566\\
25	0.015956615039315\\
30	0.0159064525276767\\
35	0.0155547442535228\\
40	0.0155564827185792\\
};
\addlegendentry{OMP with SCI}

\end{axis}

\begin{axis}[%
scale only axis,
xmin=0,
xmax=1,
ymin=0,
ymax=1,
axis line style={draw=none},
ticks=none,
axis x line*=bottom,
axis y line*=left
]
\end{axis}
\end{tikzpicture}%
}
    \caption{\ac{NMSE} comparison for six users.}
    \label{fig:CER_6_Users}
\end{figure}

Fig.~\ref{fig:CER_4_Users} and Fig.~\ref{fig:CER_6_Users} show the \ac{NMSE} vs. the pilot symbol \ac{SNR} with four and six users respectively.
Fig. \ref{fig:CER_4_Users} shows that \ac{OMP} outperforms impulse pilot-based estimation. 
However, Fig. \ref{fig:CER_6_Users} shows that impulse pilot-based estimation outperforms \ac{OMP} when we increase the number of users. From our observations, for scenarios with limited number of users as in Fig. \ref{fig:CER_4_Users}, the use of \ac{OMP} or \ac{OMP}-SCI is advisable. On the other hand, for larger number of users as in Fig. \ref{fig:CER_6_Users}, the impulse pilot-based approach proves to be a pragmatic choice. In \ac{OMP}, to maintain the performance as in the impulse pilot-based method, the number of pilots should be increased which leads to an increase in the size of $M$ and $N$. On the other hand, the impulse pilot-based approach is independent of the number of users, see Fig. \ref{fig:CER_4_Users} and Fig. \ref{fig:CER_6_Users}. 
In Fig. \ref{fig:CER_4_Users}, we can observe that by choosing the right threshold value, we can improve the estimation accuracy close to that of OMP-SCI. In both Figs. \ref{fig:CER_4_Users} and \ref{fig:CER_6_Users}, the value of threshold $\tau$ used is $3\sigma$. 

In Fig.~\ref{fig:BER_thr}, the \ac{BER} with \ac{SNR} = 10 dB for different values of threshold is shown. Evidently, the pilot estimation and data detection in the impulse pilot-based method is affected by the threshold level used for detecting the channel tap. The plot motivates the use of $3\sigma$ for the selected threshold since it shows the best \ac{BER} results in almost every scenarios considered.

The computational complexity of \ac{OMP}-SCI in terms of number of addictions and multiplications is smaller than \ac{OMP} by aligning the number of iterations with the effective number of channel taps. In general, OMP-SCI needs much smaller number of \ac{LS} operations compared to conventional \ac{OMP}. For instance, considering the scenario with four users and a fixed number of simulation cycles, it is calculated that the \ac{OMP} implies up to 35983 operations for the estimation, while the OMP-SCI reduces this number to approximately 26400 operations.

\begin{figure}[t]
    \centering
    \scalebox{0.8}{
%
%

\begin{tikzpicture}

\begin{axis}[%
scale only axis,
xmin=1,
xmax=7,
xtick={1,2,3,4,5,6,7},
xticklabels={{$\sigma$},{$\text{2}\sigma$},{$\text{3}\sigma$},{$\text{4}\sigma$},{$\text{5}\sigma$},{$\text{6}\sigma$},{$\text{7}\sigma$}},
xlabel style={font=\color{white!15!black}},
xlabel={Threshold},
ymode=log,
ymin=0.000544069640914037,
ymax=0.0309014267185473,
yminorticks=true,
ylabel style={font=\color{white!15!black}},
ylabel={BER},
axis background/.style={fill=white},
xmajorgrids,
ymajorgrids,
yminorgrids,
grid style={line width=0.3pt, draw=gray!20},
legend style={legend pos = north west, legend columns = 2, legend cell align=left, align=left, draw=white!15!black}
]
\addplot [color=blue, line width=2.0pt, mark size=1.4pt, mark=square*, mark options={solid, fill=blue, blue}]
  table[row sep=crcr]{%
1 	0.00413492927094668\\
2	0.00165941240478781\\
3	0.000897714907508161\\
4	0.00130576713819369\\
5	0.00231229597388466\\
6	0.00242110990206746\\
7	0.00552230685527748\\
};
\addlegendentry{One User Case}

\addplot [color=black, line width=2.0pt, mark size=1.4pt, mark=square*, mark options={solid, fill=black, black}]
table[row sep=crcr]{%
	1	0.00556199304750869\\
	2	0.0029837775202781\\
	3	0.00275202780996524\\
	4	0.00353418308227115\\
	5	0.00286790266512167\\
	6	0.00370799536500579\\
	7	0.00396871378910776\\
};
\addlegendentry{Two Users Case}


\addplot [color=red, line width=2.0pt, mark size=1.4pt, mark=square*, mark options={solid, fill=red, red}]
table[row sep=crcr]{%
	1	0.0135118306351183\\
	2	0.00663138231631382\\
	3	0.00482565379825654\\
	4	0.00672478206724782\\
	5	0.00853051058530511\\
	6	0.00778331257783313\\
	7	0.011239103362391\\
};
\addlegendentry{Three Users Case}

\addplot [color=green, line width=2.0pt, mark size=1.4pt, mark=square*, mark options={solid, fill=green, green}]
table[row sep=crcr]{%
	1	0.0167315175097276\\
	2	0.00885214007782101\\
	3	0.00894941634241245\\
	4	0.00843060959792477\\
	5	0.0101491569390402\\
	6	0.012905317769131\\
	7	0.015337224383917\\
};
\addlegendentry{Four Users Case}



\end{axis}

\begin{axis}[%
scale only axis,
xmin=0,
xmax=1,
ymin=0,
ymax=1,
axis line style={draw=none},
ticks=none,
axis x line*=bottom,
axis y line*=left
]
\end{axis}
\end{tikzpicture}%
}
    \caption{BER versus Threshold $\tau$ for impulse-pilot data detection.}
    \label{fig:BER_thr}
\end{figure}

\begin{figure}[t]
    \centering
    \scalebox{0.8}{
%
%
\begin{tikzpicture}

\begin{axis}[%
scale only axis,
xmin=-10,
xmax=20,
xlabel style={font=\color{white!15!black}},
xlabel={SNR [dB]},
ymode=log,
ymin=1e-05,
ymax=1,
yminorticks=true,
ylabel style={font=\color{white!15!black}},
ylabel={BER},
axis background/.style={fill=white},
xmajorgrids,
ymajorgrids,
yminorgrids,
grid style={line width=0.3pt, draw=gray!20},
legend style={legend pos=south west, nodes={scale=0.8, transform shape}, legend cell align=left, align=left, draw=white!15!black}
]
\addplot [color=blue, line width=2.0pt, mark size=1.5pt, mark=square*, mark options={solid, fill=blue, blue}]
  table[row sep=crcr]{%
-10	0.28645266594124\\
-8	0.247796517954298\\
-6	0.200489662676823\\
-4	0.156147986942329\\
-2	0.113275299238302\\
0	0.0742927094668118\\
2	0.0449945593035909\\
4	0.022714907508161\\
6	0.00992927094668117\\
8	0.00342763873775843\\
10	0.00136017410228509\\
12	0.000244831338411317\\
14	2.72034820457019e-05\\
16	0\\
18	0\\
20	0\\
};
\addlegendentry{One User Case: Imperfect CSI}

\addplot [color=blue, line width=2.0pt, dashed, mark size=1.5pt, mark=*, mark options={solid, fill=blue, blue}]
  table[row sep=crcr]{%
-10	0.28664309031556\\
-8	0.247361262241567\\
-6	0.200244831338411\\
-4	0.1560119695321\\
-2	0.112078346028292\\
0	0.0730413492927095\\
2	0.0441784548422198\\
4	0.0217083786724701\\
6	0.00889553862894451\\
8	0.0028835690968444\\
10	0.000843307943416757\\
12	5.44069640914037e-05\\
14	0\\
16	0\\
18	0\\
20	0\\
};
\addlegendentry{One User Case: Perfect CSI}

\addplot [color=black, line width=2.0pt, mark size=1.5pt, mark=square*, mark options={solid, fill=black, black}]
  table[row sep=crcr]{%
-10	0.278418308227115\\
-8	0.238470451911935\\
-6	0.194959443800695\\
-4	0.153621089223638\\
-2	0.112224797219003\\
0	0.0770857473928158\\
2	0.0469293163383546\\
4	0.0253186558516802\\
6	0.0150926998841251\\
8	0.00814020857473928\\
10	0.00460602549246813\\
12	0.00295480880648899\\
14	0.00272305909617613\\
16	0.00214368482039397\\
18	0.00170915411355736\\
20	0.00165121668597914\\
};
\addlegendentry{Two Users Case: Imperfect CSI}

\addplot [color=black, line width=2.0pt, dashed, mark size=1.5pt, mark=*, mark options={solid, fill=black, black}]
  table[row sep=crcr]{%
-10	0.278128621089224\\
-8	0.237630359212051\\
-6	0.194206257242178\\
-4	0.152230590961761\\
-2	0.111964078794902\\
0	0.0755214368482039\\
2	0.045422943221321\\
4	0.0241019698725377\\
6	0.0138760139049826\\
8	0.00692352259559676\\
10	0.0040845886442642\\
12	0.00208574739281576\\
14	0.00231749710312862\\
16	0.00168018539976825\\
18	0.00127462340672074\\
20	0.000869061413673233\\
};
\addlegendentry{Two Users Case: Perfect CSI}

\addplot [color=red, line width=2.0pt, mark size=1.5pt, mark=square*, mark options={solid, fill=red, red}]
  table[row sep=crcr]{%
-10	0.296046077210461\\
-8	0.25607098381071\\
-6	0.214788293897883\\
-4	0.171450809464508\\
-2	0.131351183063512\\
0	0.0940535491905355\\
2	0.0616749688667497\\
4	0.0397260273972603\\
6	0.0241905354919054\\
8	0.0161270236612702\\
10	0.00965130759651308\\
12	0.00638231631382316\\
14	0.00448318804483188\\
16	0.00280199252801993\\
18	0.00317559153175592\\
20	0.0024906600249066\\
};
\addlegendentry{Three Users Case: Imperfect CSI}

\addplot [color=red, line width=2.0pt, dashed, mark size=1.5pt, mark=*, mark options={solid, fill=red, red}]
  table[row sep=crcr]{%
-10	0.296170610211706\\
-8	0.255728518057285\\
-6	0.2140099626401\\
-4	0.169738480697385\\
-2	0.129825653798257\\
0	0.0923412204234122\\
2	0.0600560398505604\\
4	0.0386052303860523\\
6	0.0231631382316314\\
8	0.0144458281444583\\
10	0.00899750933997509\\
12	0.00529265255292653\\
14	0.00320672478206725\\
16	0.00180572851805729\\
18	0.00165006226650062\\
20	0.00108966376089664\\
};
\addlegendentry{Three Users Case: Perfect CSI}

\addplot [color=green, line width=2.0pt, mark size=1.5pt, mark=square*, mark options={solid, fill=green, green}]
table[row sep=crcr]{%
	-10	0.292996108949416\\
	-8	0.256225680933852\\
	-6	0.213975356679637\\
	-4	0.170330739299611\\
	-2	0.13651102464332\\
	0	0.0998054474708171\\
	2	0.0661154345006485\\
	4	0.0469844357976654\\
	6	0.0324254215304799\\
	8	0.0180285343709468\\
	10	0.0115758754863813\\
	12	0.00946822308690013\\
	14	0.00667963683527886\\
	16	0.00512321660181582\\
	18	0.00411802853437095\\
	20	0.00382619974059663\\
};
\addlegendentry{Four Users Case: Imperfect CSI}

\addplot [color=green, line width=2.0pt, dashed, mark size=1.5pt, mark=*, mark options={solid, fill=green, green}]
table[row sep=crcr]{%
	-10	0.292444876783398\\
	-8	0.256225680933852\\
	-6	0.214364461738003\\
	-4	0.170071335927367\\
	-2	0.134760051880674\\
	0	0.0990596627756161\\
	2	0.0666018158236057\\
	4	0.0458495460440986\\
	6	0.0302204928664073\\
	8	0.0165369649805447\\
	10	0.0107003891050584\\
	12	0.00736057068741894\\
	14	0.0049935149156939\\
	16	0.00291828793774319\\
	18	0.00152399481193256\\
	20	0.00100518806744488\\
};
\addlegendentry{Four Users Case: Perfect CSI}

\end{axis}

\begin{axis}[%
scale only axis,
xmin=0,
xmax=1,
ymin=0,
ymax=1,
axis line style={draw=none},
ticks=none,
axis x line*=bottom,
axis y line*=left
]
\end{axis}
\end{tikzpicture}%
}
    \caption{\ac{BER} comparison in different user scenarios.}
    \label{fig:embedded_BER}
\end{figure}
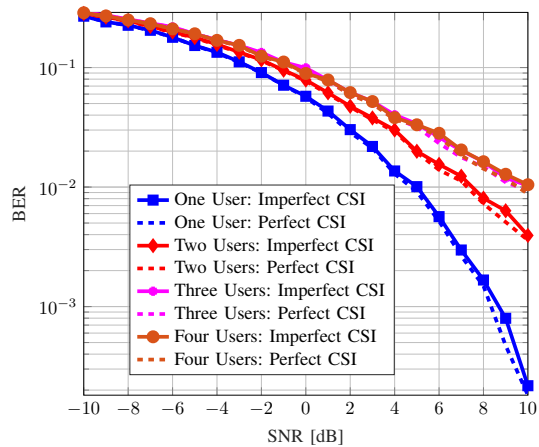

In Fig.~\ref{fig:embedded_BER}, the \ac{BER} of embedded impulse pilot-based channel estimation and data detection with the proposed pilot placement is shown. We consider four users as in Fig.~\ref{fig:embedded_pilot_four_user} and use \ac{QPSK} modulation for the data symbols. A \ac{LMMSE} decoder is used, and the results are compared to having full channel knowledge.
Fig.~\ref{fig:embedded_BER} shows the perfect and imperfect \ac{CSI} curves almost match, which demonstrates the proposed approach is reliable, especially at low \ac{SNR}. The \ac{BER} degrades depending on the number of users, but it is shown that it becomes less severe for more than three users.

\section{Conclusions}
We compared channel estimation algorithms for \ac{OTFS} in multi-user scenarios. 
Numerical results show that the choice of estimation algorithm should depend on the number of users. For instance, for a small number of users, \ac{OMP} outperforms the impulse-pilots, while for many users, impulse-pilots are better. Finally, we can lower the complexity of the conventional \ac{OMP} by considering the side channel information.

\section*{Acknowledgment}
The authors would like to thank Prof. Gerhard Kramer for constructive feedback.

The research of Lorenzo Zaniboni is funded by Deutsche Forschungsgemeinschaft (DFG) through the grant KR 3517/12-1. The research conducted by Mahdi Mahvari is funded by the 6G Future Lab Bavaria
and DFG through projects 320729232 and 509917421.

\bibliographystyle{IEEEtran}
\bibliography{radarcom,IEEEabrv}

\end{document}